# Mobile phone data reveal spatiotemporal dynamics of Omicron infections in Beijing after relaxing zero-COVID policy

Xiaorui Yan[1,2,†], Ci Song[1,2,†], Tao Pei[1,2,3,*], Erjia Ge[4], Le Liu[1,2], Xi Wang[1,2], Linfeng Jiang[5]

[1] State Key Laboratory of Resources and Environmental Information System, Institute of Geographical Sciences and Natural Resources Research, Chinese Academy of Sciences, Beijing, China

[2] University of Chinese Academy of Sciences, Beijing, China

[3] Jiangsu Center for Collaborative Innovation in Geographical Information Resource Development and Application, Nanjing, China

[4] Dalla Lana School of Public Health, University of Toronto, Toronto, Canada

[5] College of Earth and Environmental Sciences, Lanzhou University, Lanzhou, China

†These authors contributed equally to this work.
*Corresponding author: peit@lreis.ac.cn

**Funding**

This work was supported by National Natural Science Foundation of China (Grant Nos: 42071435 and 42071436) and Innovation Project of LREIS (Grant No: KPI002).

**Declaration of interest**

The authors declare no competing interests.

# Abstract

The swift relaxation of the zero-COVID policy in December 2022 led to an unprecedented surge in Omicron variant infections in China. With the suspension of mandatory testing, tracking this epidemic outbreak was challenging because infections were often underrepresented in survey and testing results, which only involved partial populations. We used large-scale mobile phone data to estimate daily infections in Beijing from November 2022 to January 2023. We demonstrated that an individual's location records of mobile phone could be used to infer his or her infectious status. Then, the derived status of millions of individuals could be summed to reconstruct the citywide spatiotemporal dynamics of infections. We found that the infection incidence peaked on 21 December, and 80.1% of populations had been infected by 14 January 2023 in Beijing. Furthermore, infection dynamics exhibited significant demographic and spatiotemporal disparities. Our work provides a ubiquitous and high-coverage data source for monitoring epidemic outbreaks.

# Main text

At the end of 2022, China began to relax its dynamic zero-COVID policy, which was maintained for more than two years[1]. Most stringent control measures were abandoned, including large-scale lockdowns, mandatory nucleic acid testing, and centralized isolation, after the announcements of the 20 measures on 11 November 2022[2] and 10 measures on 7 December 2022[3]. The rapid relaxation of restrictions led to a spike in severe acute respiratory syndrome coronavirus 2 (SARS-CoV-2) Omicron lineage BA5.2 and BF.7 infections in many cities, including Beijing[4,5]. Reconstructing the spatiotemporal dynamics of infections contributes to understanding the process by which a pandemic with high infectivity strikes a city, and therefore is crucial for decision-making in disease prevention and control. However, with the suspension of regular mass testing and intensive contact tracing, official daily case data were no longer reliable after the policy adjustment[3], leading to the difficulty in tracking the infection dynamics.

Until now, the daily infections of this epidemic wave were usually estimated using modelling with survey data[6-9], testing results[10-13], and some auxiliary information[6,14]. The reconstructed infection dynamics might deviate from the truth, which was predominantly caused by data biases. For instance, a widely used source of data, online surveys from social media, such as Sina Weibo and Tencent WeChat, tracked a small number of individuals (mostly younger people) at infrequent intervals[6,7]. Furthermore, self-reported infection status might also be unreliable and therefore bias the survey results[8,9]. Testing results, including nucleic acid tests and serological tests, were limited to partial populations because only patients with severe symptoms and elderly individuals tended to take those tests after the abandonment of regular mass testing[10-13]. Auxiliary information was mainly composed of crowd mobility data derived from subway passengers[6] and comparative data from other countries[14]. However, the former might lack the representation of the population that uses other transportation means, and the latter might be biased because of the differences among countries, such as in vaccine uptake.

Mobile phone data, as a kind of reliable data source associated with individual behavior of large-scale populations, have been commonly used in public health studies[15-17]. Here, we used daily mobile phone data over a 2-month period (14 November 2022 to 14 January 2023) covering ~12 million unique subscribers in Beijing, equivalent to ~60% of the total population[18], to capture individuals' mobility status, that is, staying home or going outside for work, recreation, etc. Using changes in daily mobility status, we then tracked individual infection status and finally reconstructed the citywide dynamics of infections after the relaxation of the zero-COVID policy with the aim of estimating the incidence and cumulative attack rate (i.e., proportion of the population infected since 14 November 2022), as well as capturing the spatiotemporal spread of infections.

## Results

**Dynamics of the incidence and cumulative attack rate**

With the rapid outbreak of Omicron BF.7 in Beijing, many symptomatic infected individuals self-isolated at home for treatment and recovery after the relaxation of control measures[3]. Using daily anonymized mobile phone data from 14 November 2022 to 14 January 2023, we inferred individual infectious status from daily mobility status, that is, staying home or going outside (**Fig. 1a**) and then reconstructed citywide dynamics of infections. See **Methods** section for details.

**Fig. 1b** presents the citywide epidemic curves for the incidence and cumulative attack rate of infections based on individuals' onset dates. We found that the daily incidence formed a bell-shaped curve, experiencing a peak after the easing of the zero-COVID policy. On 29 November 2022, we estimated that the incidence of infections was 0.42% (95% credible interval (CI): 0.25–0.55). After mass testing, intensive contact tracing and lockdown measures were completely suspended on 30 November 2022, the incidence rapidly increased and peaked on 21 December 2022, with an estimated incidence of 2.69% (95% CI: 2.05–2.89) and cumulative attack rate of 42.83% (95% CI: 27.67–50.45). As the susceptible population decreased and more people recovered from infections with antibodies, the daily incidence started to decrease and returned to a low level in early January 2023. By 14 January 2023, we estimated that the cumulative attack rate was 80.1% (95% CI: 58.84–90.42), and this wave of infections ceased with an estimated incidence of 0.008% (95% CI: 0.007–0.01).

We then analyzed the infection dynamics for different age groups, namely, 0–25 years, 26–60 years, and ≥ 61 years. As the Omicron variant demonstrated high transmissibility and immune escape capability[4,19], all age groups were heavily affected. By 14 January 2023, we estimated that the cumulative attack rates were 79.88% (95% CI: 56.38–90.35), 79.3% (95% CI: 57.01–90.08), and 82.06% (95% CI: 63.53–91.24) for those aged 25 years and under, 26 to 60 years, and ≥ 61 years, respectively (**Fig. 1c**). We also found that the daily incidence for each group showed a bell-shaped curve and experienced a peak on approximately 21 December 2022 (**Supplementary Fig. 1**). While the final cumulative attack rates and shapes of daily incidence were similar, the incidence still demonstrated several different temporal patterns among different age groups (**Fig. 1c**). Specifically, the incidence among those aged 25 years and under significantly exceeded those of other age groups before 20 December 2022 due to their higher mobility and possibility of exposure to the virus after the end of the control measures. In contrast, we found a higher incidence among older people during the late phase of the epidemic.

**Spatiotemporal spread of infections**

To track the spatiotemporal disparities in Omicron transmission in Beijing, we aggregated inferred symptomatic individuals at the district level according to their home locations. We found that the estimated incidence formed a bell-shaped curve for all districts and peaked on approximately 21 December 2022 (**Supplementary Fig. 2**). We classified districts into three groups based on the changes in the daily incidence (normalized to the difference from the average) using a clustering approach (**Fig. 2a**). Group 1 (red) experienced a "high–low" shape of the incidence curve, that is, the incidence was the highest among all districts during the initial phase of this epidemic wave and returned to the lowest during the later phase. In contrast, group 2 (blue) experienced a "low–high" shape of the incidence curve with a lower incidence in the initial phase and a higher incidence in the later phase. The incidence in group 3 moderately fluctuated above the average level throughout the whole epidemic. We mapped the three groups of districts and found that the dynamics of infections were heavily dependent on geographical location, as shown in **Fig. 2b**. Districts in group 1 were mainly in the central areas of Beijing with high population density and mobility, while districts in group 2 were concentrated in sparsely populated suburbs. Districts in group 3 were scattered in the transition zone between these areas.

We also estimated the cumulative attack rate for each district (**Fig. 2c** and **Supplementary Fig. 2**). As of 14 January 2023, this wave of infection affected the Dongcheng, Haidian, Shijingshan, and Fangshan districts, which are mostly in the city center, the most heavily, while the Pinggu, Miyun, Huairou, and Tongzhou districts, which are mainly in the suburbs, had lower cumulative attack rates. Combined with the precise location information provided by mobile phone signaling data, our method provided the geographical distribution of infections at an extremely fine degree of spatial granularity because the mobile phone data enabled us to track the transmission of infections at the 500 m × 500 m grid level (**Fig. 2d**). Global spatial autocorrelation (Moran's $I$ = 0.13, $p$ value < 0.001, see **Supplementary Methods**) demonstrated that the grid-level cumulative attack rate was spatially aggregated. We further identified local clusters of the cumulative attack rate using a local Moran's statistic (**Supplementary Methods**). As illustrated in **Supplementary Fig. 4**, areas with High-High significant spatial autocorrelation were mainly located within the Fifth Ring Road, where the vast majority of Beijing's population and urban facilities are, and in some large residential communities in the suburbs, e.g., Huilongguan and Tiantongyuan. Low-Low clusters were mainly found in the eastern border regions with sparse populations and less residents' mobility.

**Comparison with previous estimates**

To validate our results, we collected three sources of estimates for this wave of Omicron infections: 1) the daily number of infections estimated in a recent study conducted by Leung et al.[6], from which we only selected the data during our study period for comparison; 2) the test positivity rate reported by the Chinese Center for Disease Control and Prevention (China CDC, https://en.chinacdc.cn/) from 9 December

2022[20]; and 3) the proportion of participants self-reported to be infected from 10 December to 28 December 2022 in Beijing, based on YiLuomu's Weibo online survey (https://weibo.com/2987102112/MlFnEt1zf). To reveal the difference in dynamic patterns and peak time more clearly, we used the Min-Max normalization method (see **Methods** for details) to transform each dataset and our estimated daily incidence into the same range (i.e., [0, 1]), as shown in **Fig. 3**.

Our epidemic curve showed a similar shape to those of the data from the China CDC and YiLuomu's online survey, that is, a bell shape with a peak. In contrast, Leung et al.[6] estimated that the incidence peaked twice in Beijing (on 10 December and 21 December 2022). Our estimation peaked on 21 December 2022, which was a compromise of YiLuomu's survey results (17 December 2022) and that of the China CDC (25 December 2022) and consistent with the second peak from Leung et al.[21] The reasons for the differences were due to the data's origins. On the one hand, most active Weibo users were younger people, those with higher mobility and risk of exposure to viruses; as a result, the number of infections from YiLuomu's survey peaked earlier. On the other hand, the peak from the China CDC appeared later because Beijing was in the first group of cities with a surge of infections in China. The comparison demonstrated the reasonability of the shape and peak of our result.

**Discussion**

The National Health Commission of China (http://www.nhc.gov.cn/) stopped announcing daily COVID-19 cases after the end of regular mass testing. Regarding the surge of infections, previous works tracked dynamics of this epidemic wave based on sampling data covering partial populations, such as online surveys[6,7,9] and auxiliary information, e.g., crowd mobility derived from subway passengers[6]. This study is the first to estimate infection incidence with data from ubiquitous mobile phone data. In this context, we captured individual mobility behaviours and used changes in mobility patterns to fully reconstruct the epidemic dynamics of the Omicron outbreak for the entire Beijing after the abandonment of the zero-COVID policy. We found that this wave of the Omicron epidemic peaked on 21 December 2022 in Beijing, and 80.1% of the individuals had been infected by 14 January 2023. After relaxing the control measures, the infections spread faster among younger people during the initial phase because of their higher mobility, but the daily incidence was higher in the older age group during the later phase. The transmission process showed significant spatiotemporal heterogeneity in Beijing. The daily incidence in central areas exceeded that in suburbs during the initial phase of the epidemic, and the reverse occurred later. Furthermore, this epidemic wave had the largest impact on the city center, which had the highest cumulative attack rate.

The major advantage of this study is that because of the ubiquity of mobile phone and detailed location information provided by it, we could capture the epidemic spread of a large-scale population (including people of different ages) over a long-term period (two months before, during, and after this epidemic wave) with fine-grained spatial resolution (i.e., at the hundred-meter level)[15,22]. While a few studies have modelled transmission using mobile phone data for previous epidemic waves[23,24], they integrated the digital proxies of population mobility derived from aggregate data into conventional epidemic models. Our study tracked individuals' infectious status directly from individual-level data, providing a higher temporal and spatial resolution of infection dynamics. In the meanwhile, the individual-level data were by default anonymized and therefore should not elicit privacy concerns. An additional advantage of our approach is its generalizability: using mobile phone data, the methodology can be easily generalized to other cities or countries and applied to other contagious disease pandemics in the future because of the wide use of mobile phones.

Our study has several limitations. First, while mobile phone data have improved the coverage of populations and time periods, issues related to the representativeness of the data still exist. Specifically, our data might not include some vulnerable populations, such as children and lower-income individuals who are less likely to own a phone[25]. As such, the representativeness of our results could be enhanced by including data that record more specific information for these populations (e.g., smartwatches for kids) in future studies. Second, our analysis was limited to symptomatic individuals, and asymptomatic individuals might be excluded because they did not experience a long-stay-home period. However, symptomatic infections were surveyed to account for over 95% of cases[26]. Thus, the impacts of such bias may not significantly affect the estimated results. Third, our method can only be applied under two conditions: control measures such as lockdowns are suspended, and the virus is not too strongly pathogenic that it frightens people into staying at home to prevent infections.

# Methods

## Data

### Mobile phone signaling data

We collected anonymous mobile phone signaling data from one of Beijing's three communication companies. The data contained daily location information of over 12 million mobile phone subscribers, accounting for more than 60% of the population in Beijing[18], from 14 November 2022 to 14 January 2023. Specifically, each record included the unique identifier for the subscriber, the date and time of this record, and the location of the cell tower to which the phone was connected. An example of a subscriber's mobile phone records during one day is shown in **Supplementary Table 1**. We used the locations of cell towers to approximately represent the subscribers' locations. The subscribers' unique identifiers were fully encrypted to protect personal privacy. In addition to location information, the collected data also contained subscribers' attributes, including age group and sex.

### Reconstructing the dynamics of Omicron infections

As the majority of patients in Beijing chose to self-isolate at home for treatment and recovery after the relaxation of the zero-COVID policy[21], we derived the individual infectious period from the stay-at-home period and subsequently tracked the dynamics of infections within the whole city. Our study period was from 14 November 2022 to 14 January 2023, covering the periods before, during, and after this wave of infections. Using large-scale mobile phone data, we developed a four-step methodology to reconstruct the infection transmission process.

*Step 1: Extracting individual daily movement trajectory.* Based on mobile phone signaling data, we extracted the daily movement trajectory of each subscriber by sorting locations in ascending order of the recording time. To further obscure subscribers' exact locations and reduce computing costs, trajectory points were aggregated to a grid (the study area, Beijing, was divided into a set of 500 m × 500 m regular grids), and each hour was assigned a grid where the subscriber stayed the longest during this hour. In this manner, we obtained the hourly grid trajectory of a subscriber, e.g., $trj = \{(grid_1, 1), (grid_2, 2), \ldots, (grid_t, t), \ldots, (grid_{24}, 24)\}$, where $grid_t$ denotes the grid with the longest stay during hour $t$. We filtered out the subscribers who missed location information records for over 12 hours each day from the analysis.

*Step 2: Identifying subscribers' home locations.* Using the extracted daily movement trajectory, we first identified the location where a subscriber stayed the longest during night-time hours (22:00–06:00) as the candidate home location for each day. Then, we counted the number of days that each candidate home location appeared during the study period and selected the most frequent one or two (if more than one existed) as the final home location. We used a 500 m × 500 m regular grid as the areal unit to preserve privacy, and the home location for each subscriber was indicated by a grid cell. Specifically, given that the hourly grid trajectory on day $d$ of subscriber $u$ is $trj_{u,d} = \{(grid_1, 1), (grid_2, 2), \ldots, (grid_t, t), \ldots, (grid_{24}, 24)\}$, the home location of subscriber $u$ was identified by the argmax function as follows.

$$H_{u,d} = \mathrm{argmax}_{grid \in G_{u,d}} \left( Num(grid | t \in T_{night}) \right) \qquad (1)$$

$$H_u = \mathrm{argmax}_{grid \in \forall H_{u,d}} (Num(H_{u,d})) \qquad (2)$$

Where $H_{u,d}$ is the candidate home location derived on day $d$ of subscriber $u$, $H_u$ is the final home location of subscriber $u$, $G_{u,d}$ is the set of all grid locations by subscriber $u$ on day $d$, $Num(\cdot)$ is a count function that returns the number of hours that the subscriber stayed at $grid$ in equation (1) or the number of days that the candidate home location $H_{u,d}$ appeared during the whole study period in equation (2), and $T_{night}$ is the set of night-time hours (22:00–06:00).

*Step 3: Selecting the long-stay-home period for recovery from infection.* The long-stay-home period was defined as a sequence of several continuous days during which a person stayed at home. Patients usually experienced the long-stay-home period for self-treatment and recovery. Based on the subscribers' daily grid trajectories derived from step 1 and grid-based home locations derived from step 2, we first determined whether a subscriber stayed home for a given day using the formula

$$S_{u,d} = \begin{cases} 0, if\ |\{grid_t | grid_t \in G_{u,d}, grid_t \neq H_u\}| > 1 \\ 1, otherwise \end{cases} \quad (3)$$

where $S_{u,d}$ is an indicator of stay-at-home status for subscriber $u$ on day $d$, specifically, $S_{u,d} = 1$ indicates that the subscriber stayed at home on day $d$, and $S_{u,d} = 0$ indicates that the subscriber went out for work, recreation, etc. In other words, if one subscriber was not at home for no more than 1 hour on one day, he or she was assumed to be staying at home on that day.

During the study period (covering 61 days), the daily stay-at-home status formed a time series for each subscriber, given by

$$HS_u = \{S_{u,1}, S_{u,2}, \dots, S_{u,d}, \dots, S_{u,61}\} \quad (4)$$

Thus, the daily stay-at-home status of each subscriber was converted into a 61-day-long time series composed of 0 or 1. Based on the derived time series, we assumed that if a subscriber was infected with symptoms, he or she would stay at home for more than a given number of days (how to set the threshold will be introduced later) continuously, which was referred to as the long-stay-home period. However, since people might stay home for other purposes other than infection, such as working from home[27], the series of a subscriber was likely to contain multiple long-stay-home periods; for instance, a subscriber with $HS_u = \{0,0,1,1,1,1,0,\dots,0,1,1,1,1,1,1,0\}$ had two long-stay-home periods if the threshold was set to 3, i.e., $\{1,1,1,1\}$ and $\{1,1,1,1,1,1\}$. As a result, with the assumption above, we determined which long-stay-home period was the one during which the subscriber was infected with symptoms, and the procedure was as follows (see the flow chart in **Supplementary Fig. 5**). If the ending date of the last long-stay-home period was before 7 December 2022, then the longest of those was thought of as the one during which the subscriber was infected. Otherwise, there were one or more long-stay-home periods that ended later than 7 December 2022; in such cases, the longest among those was determined as the one during which the subscriber was infected (see examples in **Supplementary Fig. 5**).

The reason for setting 7 December 2022 as the boundary to differentiate the determination was because the new 10 measures were announced on that date[3]. The 10 measures fully ended lockdown measures, encouraged residents to resume work and schools to function normally, and allowed those infected to undergo home quarantine for self-treatment[3]. Therefore, residents tended to go outside but for being infected after 7 December 2022. Furthermore, we assumed that infected subscribers needed more time to stay home for recovery compared with noninfected subscribers. If all long-stay-home periods of a subscriber occurred before 7 December 2022, indicating that he or she was infected before that date, we then chose the longest period among those as the one during which the subscriber was infected. After the lockdown ended on 7 December 2022, people were not required to stay at home[3]. As a result, if there were still one or more long-stay-home periods after that date, the longest among them might mean that the subscriber was infected. In other words, we chose the longest period from those long-stay-home periods after 7 December even if there were longer long-stay-home periods before that date.

*Step 4: Inferring individual's symptom onset time and citywide epidemic curve.* We assumed that symptom onset occurred during the long-stay-home period derived from *Step 3,* and the end of staying home represented the time when symptoms disappeared; that is, patients went out once they recovered. Under this assumption, we utilized Bayesian inference[28] to infer the patient's symptom onset time by incorporating the probability distribution of symptom duration. Given that the derived long-stay-home period of a patient started on the $i$th date and ended (symptoms disappeared) on the $j$th date, the probability of symptom onset on the $n$th date ($i \leq n < j$) was calculated as follows:

$$P(O_n|D_j) = \frac{P(D_j|O_n)P(O_n)}{P(D_j)} = \frac{P(D_j|O_n)P(O_n)}{\sum_{n=i}^{j-1} P(D_j|O_n)P(O_n)} \quad (5)$$

where $P(O_n|D_j)$ is the posterior probability of symptom onset on the $n$th date ($O_n$) on the condition that symptoms disappeared on the $j$th date ($D_j$), $P(D_j|O_n)$ is the likelihood of symptom onset on the $n$th date given that symptoms disappeared on the $j$th date, and $P(O_n)$ and $P(D_j)$ are the prior probabilities. $P(O_n)$ was assumed to be constant, i.e., $P(O_i) = P(O_{i+1}) = \cdots = P(O_{j-1})$; thus, $P(O_n|D_j) = \frac{P(D_j|O_n)}{\sum_{n=i}^{j-1} P(D_j|O_n)}$.

The likelihood ($P(D_j|O_i)$) was estimated by fitting a Weibull distribution[29] to survey data on symptom duration of recovered patients, which was conducted via the RenSheTong online platform, provided by the Chinese Ministry of Human Resources and Social Security on 26 December 2022 (https://wj.qq.com/s2/11398030/0ffd/). See the survey data and fitting results in **Supplementary Fig. 6**. Based on the individual's distribution of symptom onset time, we obtained the citywide epidemic curve for all patients by statistical analysis.

The threshold for the number of days for defining the long-stay-home period in *Step 3* was set according to our recent online survey (https://wj.qq.com/s2/11990736/856b/) on the duration of the long-stay-home period during infection (data from the survey are shown in **Supplementary Fig. 7**). Specifically, we fitted a log-normal distribution to the survey data. The thresholds were set to 4, 7, and 15 days, corresponding to 2.5%, 50%, and 97.5% of participants in the fitting result, respectively. In this manner, we obtained the 95% credible intervals and the central line of the epidemic curve.

**Comparison and validation using previous estimates**

We used three sources of estimates for this wave of Omicron infections to validate our estimated daily incidence. The first was the estimation results from a recent study using a conventional epidemic model that tracked the dynamics of the same epidemic in Beijing from 1 November 2022[6], in which the daily number of infections and cumulative attack rates were estimated. We used the daily number of infections for comparison. Second, we collected China CDC reports on 8 January 2023[20], including the polymerase chain reaction (PCR) and antigen test results, outpatient and inpatient data, and vaccination rate after 9 December 2022. We used daily PCR positivity rates for comparison. Third, a Weibo online poll (https://weibo.com/2987102112/MlFnEt1zf)

conducted by user YiLuomu provided survey data on the participants' infection status ("uninfected", "infected", and "recovered") from 10 December to 28 December 2022, covering all provinces in China. In comparison with our estimates, we selected the data of the daily proportion of participants self-reported to be infected in Beijing.

To show the difference in dynamic patterns and peak times more clearly, the Min-Max normalization method[30] was utilized to transform each source of estimates into the same range of [0, 1]. Taking the daily incidence estimated in this study as an example, its normalization was performed as follows:

$$Norm\_inc_i = \frac{Inc_i - \min(Inc)}{\max(Inc) - \min(Inc)} \qquad (6)$$

where $Norm\_inc_i$ is the normalized incidence on date $i$, $Inc_i$ is the original incidence on date $i$, and $\min(Inc)$ and $\max(Inc)$ are the minimal and maximal daily incidence during the study period, respectively.

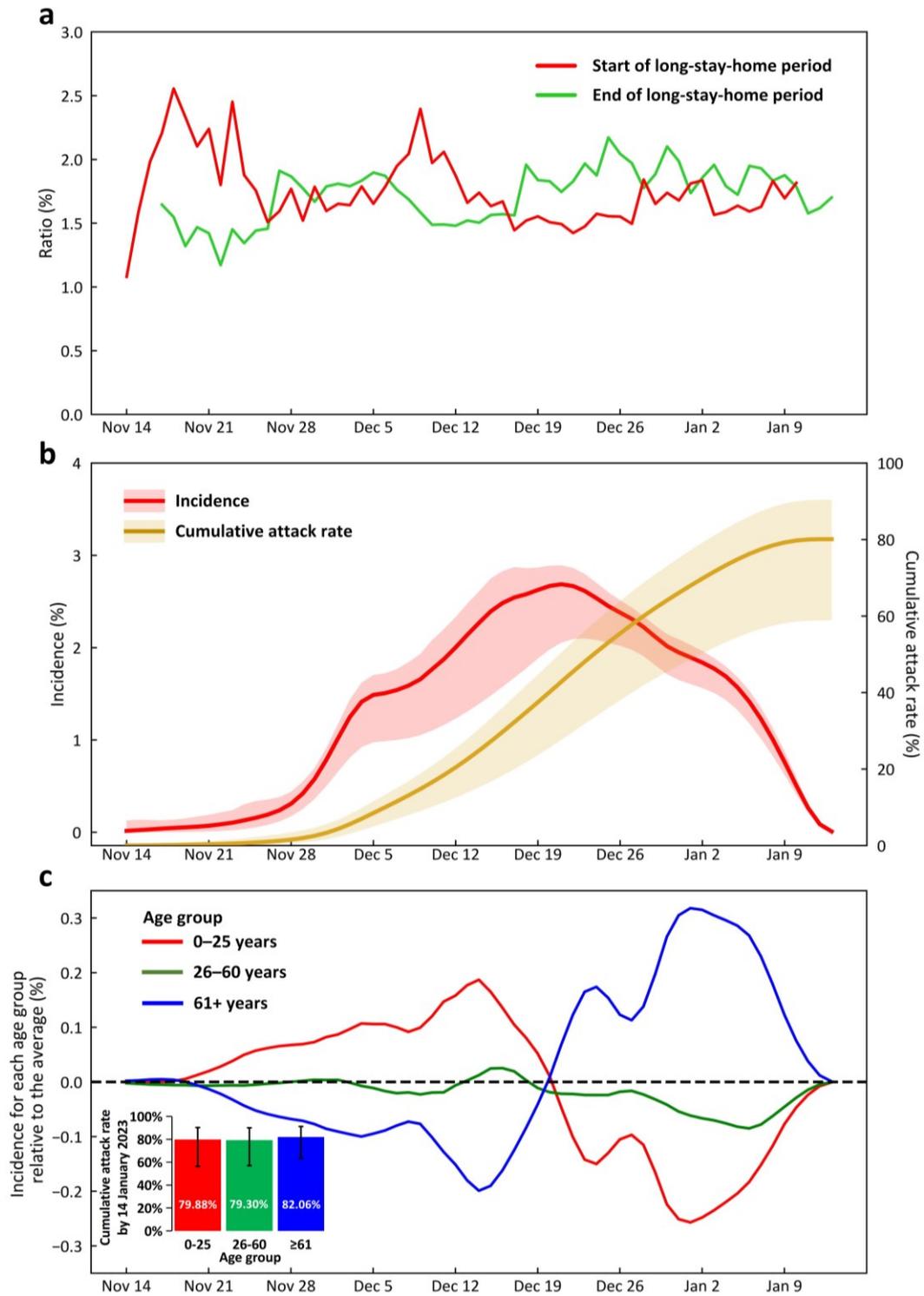

**Fig. 1. Daily incidence and cumulative attack rate of infections based on symptom onset times in Beijing. a**, The ratios of individuals who started their long-stay-home periods and those who ended their long-stay-home periods. **b**, Estimated daily incidence and cumulative attack rate. Shaded areas indicate 95% credible intervals. **c**, Estimated incidence for each age group relative to their average, that is, the difference from the average. Bars represent the cumulative attack rate for each age group by 14 January 2023.

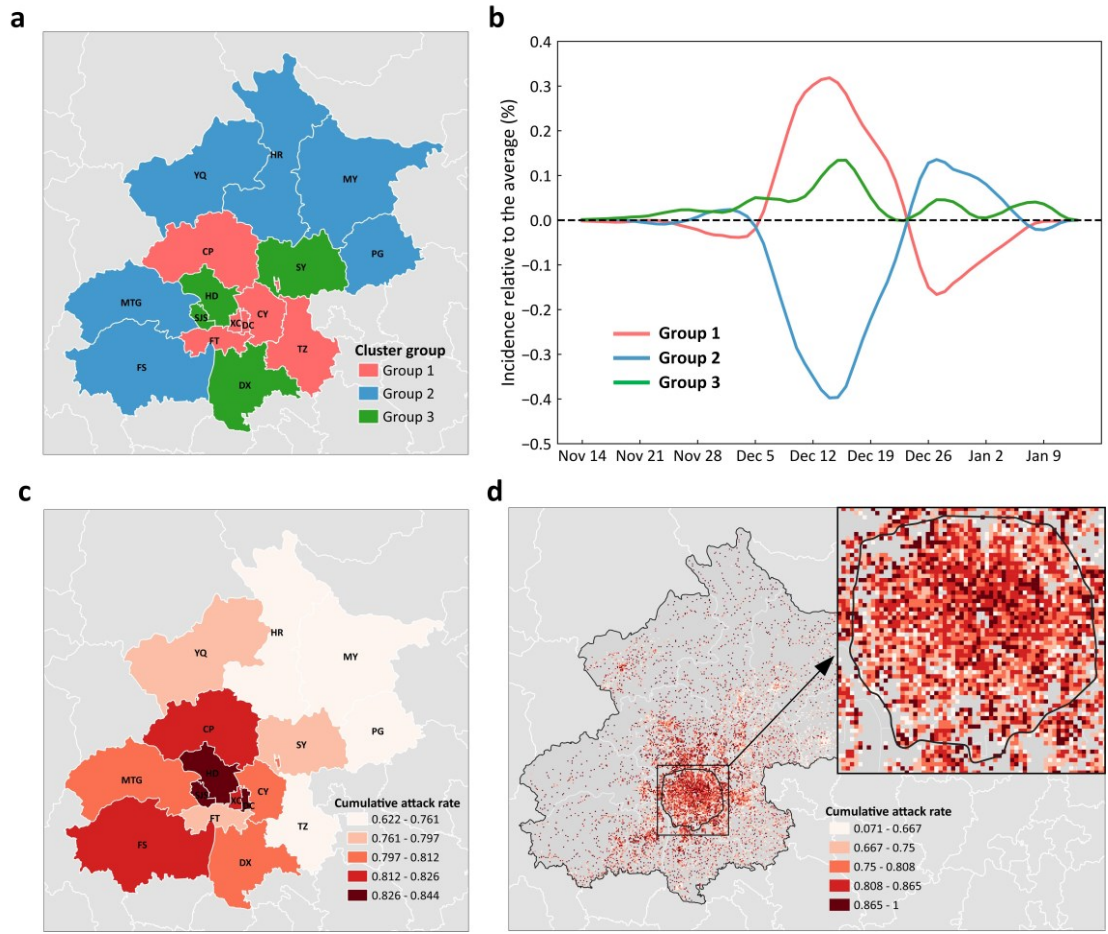

**Fig. 2. Spatiotemporal spread of Omicron infections in Beijing. a**, Geographical map of each group divided by a clustering approach based on the dynamics of the daily incidence, normalized to the difference from the average. **b**, Cluster center of each group of time series. **c**, Estimated cumulative attack rate for each district by 14 January 2023. **d**, Estimated cumulative attack rate for each 500 m × 500 m grid by 14 January 2023. The top right inset rectangle displays the enlargement of the region within the Fifth Ring Road.

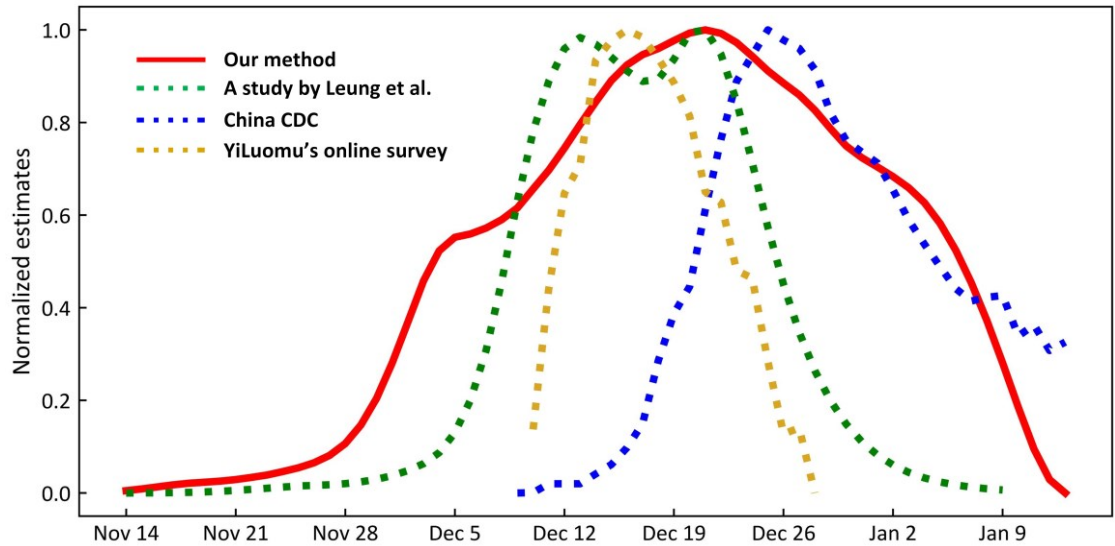

**Fig. 3. Comparison between our estimates with three datasets.** The red solid line represents our estimated daily incidence after normalization. The green, blue, and yellow dotted lines represent the normalized daily number of infections from the study by Leung et al.[6], the test positivity rate reported by the China CDC, and the proportion of participants self-reported to be infected in Beijing from YiLuomu's Weibo online survey, respectively.